\documentclass[a4paper,10pt]{scrartcl}
\usepackage[utf8]{inputenc}

\usepackage{natbib}
\usepackage{url}

\usepackage{graphicx}
\usepackage[section]{placeins}
\usepackage{verbatim}

\usepackage{lineno}

\makeatletter
\def\verbatim@font{\footnotesize\ttfamily}
\makeatother

\title{Animal Movement Tools (\texttt{amt}): R-Package for Managing Tracking Data and Conducting Habitat Selection Analyses}
\author{Johannes Signer, John Fieberg \& Tal Avgar}
\date{}

\begin{document}

\maketitle

\section*{Author's Affiliations}

\begin{itemize}
\item Johannes Signer\footnote{\texttt{jsigner@gwdg.de}}: Wildlife Sciences, University of Goettingen, Göttingen, Germany
\item John Fieberg: Department of Fisheries, Wildlife and Conservation Biology, University of Minnesota, St. Paul, MN, USA
\item Tal Avgar: Department of Integrative Biology, University of Guelph, Guelph, ON, Canada
\end{itemize}

\newpage

\section*{Summary}

\begin{enumerate}
\item Advances in tracking technology have led to an exponential increase in animal location data, greatly enhancing our ability to address interesting questions in movement ecology, but also presenting new challenges related to data management and analysis.
\item Step-Selection Functions (SSFs) are commonly used to link environmental covariates to animal location data collected at fine temporal resolution. SSFs are estimated by comparing observed steps connecting successive animal locations to random steps, using a likelihood equivalent of a Cox proportional hazards model. By using common statistical distributions to model step length and turn angle distributions, and including habitat- and movement-related covariates (functions of distances between points, angular deviations), it is possible to make inference regarding habitat selection and movement processes, or to control one process while investigating the other. The fitted model can also be used to estimate utilization distributions and mechanistic home ranges.
\item Here, we present the R-package \texttt{amt} (animal movement tools) that allows users to fit SSFs to data and to simulate space use of animals from fitted models. The \texttt{amt} package also provides tools for managing telemetry data.
\item Using fisher (\textit{Pekania pennanti}) data as a case study, we illustrate a four-step approach to the analysis of animal movement data, consisting of data management, exploratory data analysis, fitting of models, and simulating from fitted models.
\end{enumerate}

\newpage

\section*{Introduction}

Advances in technology have led to large collections of fine-scale animal biotelemetry data \citep{cagnacci2010animal, kays2015terrestrial}, fueling the development of new quantitative methods for studying animal movement \citep{hooten2017animal}. \cite{nathan2008movement} introduced the movement ecology paradigm, that conceptually connects different factors shaping the realized movement path of animals (e.g., the internal state of an animal, interaction with intra- and conspecifics, and varying environmental conditions). The movement ecology paradigm can serve as a framework for generating new hypotheses about animal movements. To test these hypotheses, efficient and straightforward tools for the management and analyses of movement data are required. Although a large number of R packages have been developed for analyzing animal movement data \citep[e.g.,][]{calabrese2016ctmm, gurarie2009novel, michelot2016movehmm} these packages often utilize domain-specific data formats and focus on a narrow subset of analytical methods (e.g., methods for fitting discrete or continuous time movement models, trajectory segmentation). We had two primary objectives in developing the \texttt{amt} R-package, namely to provide: 1) a set of functions for exploratory analyses of movement data in R, and 2) functions that facilitate the analysis of fine-scale animal location data using Step-Selection Functions (SSFs). Step-Selection Functions are powerful tools for modeling animal movement and habitat selection, but are not currently available in open-source software packages, despite their popularity.

Methods that quantify habitat selection by linking environmental covariates to location data of animals have been around for a long time. Traditionally Resource Selection Functions \citep[RSF;][]{boyce1999relating, manly2007resource} were used to study habitat selection of animals. RSFs compare covariates associated with locations where the animal was observed with covariates associated with random locations within the 'availability domain', a spatial domain within which any location is assumed available for the animal to use at any given time. Despite the sensitivity of the resulting inference to habitat availability \citep{beyer2010interpretation}, no consensus exists as to the most suitable approach to delineate the spatial domain of availability \citep{northrup2013practical,paton2016defining,prokopenko2017extent}. Moreover, the assumption that the availability domain can be considered temporally static might have been justifiable for very coarse sampling rates (e.g., daily or weekly positions of the animal), but is challenging for modern GPS data with sampling rates $<$ 1 hour.  Step-Selection Functions \citep[SSFs;][]{fortin2005wolves, thurfjell2014applications} resolve these issues by pairing each observed location with a set of random locations deemed accessible from the previously observed location. Step-Selection Functions estimate conditional selection coefficients using a likelihood equivalent of a Cox proportional hazards model \citep{gail1981likelihood}.

Until recently, SSFs were fitted by sampling random points based on the empirical (observed) distribution of 'steps' (straight lines connecting consecutive locations). This approach has come under some scrutiny as it implicitly assumes habitat selection is conditional on animal movement but not vice versa, potentially leading to biased inference \citep{forester2009accounting}. A recent extension, termed integrated SSF (iSSF), alleviates this concern and allows for simultaneous inference of habitat selection and movement processes \citep{avgar2016}. This is accomplished by requiring that random steps are sampled under one of several analytical distributions, and also by including, in addition to habitat-related covariates, movement-related covariates (functions of distances between points, angular deviations) resulting in likelihood-based estimates of the shape and scale of the underlying analytical distributions \citep{avgar2016, duchesne2015equivalence, forester2009accounting}. Unlike SSFs (that do not include an explicit movement component), a fitted iSSF is a fully-fledged biased random walk model that can be used to simulate animal space-use \citep{duchesne2015equivalence, avgar2016, signer2017}. 

Step-selection functions (SSFs and iSSFs) are usually straight forward to fit (using any conditional-logistic regression routine) once data are appropriately structured, but data preparation itself tends to be more complex and confusing and may thus become a limiting step in the application of this approach. Here, we describe the \texttt{amt} package for R, which provides a flexible and coherent workflow for efficient analysis of animal tracking data. We make heavy use of piped workflows and list-columns as introduced to R through the \texttt{tidyverse} package-family \citep{tidyverse}. We illustrate a typical workflow for fitting a (i)SSF using fisher (\textit{Pekania pennanti}) data from \cite{lapoint2013animal}. Detailed vignettes, help files, sample data and analyses are available in the amt package available on CRAN (\url{https://cran.r-project.org/package=amt})

\section*{Functionality}

A typical workflow to analyze animal tracking data can be divided into four main steps (described in detail below):

\begin{enumerate}
\item Data preparation, inspection and management: Load and inspect gaps in the data, resample tracks if needed, and adjust coordinate reference systems. 
\item Exploratory data analysis and descriptive analyses: Explore patterns in the data graphically, consider multiple movement characteristics (e.g., step-length distribution, net square displacement, or home-range size) across several animals and/or time periods.
\item Modeling: Fit models to answer questions or test hypothesis related to movement and space use of animals.
\item Simulation: Use fitted models to simulate derived quantities (e.g., space use) and assess model fit.
\end{enumerate}

\paragraph{Data preparation, inspection, and management}

After loading data into R, users should perform a variety of data quality checks and possibly remove fixes with missing coordinates (although this information could potentially be used to test if fixes are missing at random). We provide functions to quantify variability in sampling rates over time and among individuals, inspect the data visually for obvious outliers (e.g., determined by screening for unreasonable speeds), remove periods at the beginning and the end of the track to exclude possible capture effects, and resample the data to form regular bursts (i.e., partition the track into groups of observations with regular sampling rates, within some specified level of tolerance).

\paragraph{Exploratory data analysis and descriptive analyses}
Once data have been cleaned, the next logical step is to explore the data by looking at different movement-related statistics (e.g., distributions of turning angles or step lengths) and trajectory and space-use summaries (e.g., net squared displacement, path sinuosity, home range area). These summaries may be calculated for the whole trajectory or on a subset of points (a track might be split by time of the day, season, year or any other biologically meaningful factor).

\paragraph{Modeling}
In the next step, we fit models to test hypotheses about animal movement and habitat selection. Importantly, \texttt{amt} provides functionality necessary for data development steps prior to fitting RSFs and (i)SSFs (e.g., methods for generating random points or steps, and extract environmental covariates for the observed and random steps). For many other analyses (e.g., behavioral change point analyses, fitting continuous time movement models or identification of hidden behavioral states with hidden Markov models), \texttt{amt} provides coercion functions to translate location data into objects of classes required by the respective packages.

\paragraph{Simulation}
As a final step, new data can be simulated from fitted models. Simulations can be used to obtain estimates of space use (i.e. the utilization distribution), identify corridors of high use, or asses the power of the model (testing how well parameters can be recovered as a function of sample size). Many packages that fit models also provide methods to simulate from fitted models (e.g., \texttt{ctmm} or \texttt{moveHMM}). \texttt{amt} provides means to simulate space use from fitted SSFs.

\section*{Case study}
We illustrate a subset of the above steps using data from radio collard fishers available through movebank \citep{lapoint2013animal,lapoint2013data}. For details about the data and the capture of the animals, we refer to \cite{brown2012accelerometer} and \citet{lapoint2013animal}. We begin by analyzing the space use of Ricky T (id 1016), and then illustrate how similar analyses can be extended to several animals for population-level inference \citep{fieberg2010correlation}.

\subsection*{From data cleaning to simulated space use for one animal}
We begin with loading the data of all fishers, remove observations with missing spatial coordinates (longitude, latitude), and subset relocations for Ricky T (id: 1016).

\begin{verbatim}
library(raster)
library(lubridate)
library(amt) 
dat <- read_csv("data/Martes pennanti LaPoint New York.csv") %>% 
  filter(!is.na(`location-lat`)) %>% 
  select(x = `location-long`, y = `location-lat`, 
         t = `timestamp`, id = `tag-local-identifier`) %>% 
  filter(id %in% c(1465, 1466, 1072, 1078, 1016, 1469))
dat_1 <- dat %>% filter(id == 1016)
\end{verbatim}

The function \texttt{mk\_track} creates a track (the basic building block of the \texttt{amt} package), given the names of the columns containing \texttt{x} and \texttt{y} coordinates, time (\texttt{t}), and we can set a coordinate reference system (CRS). The original data was provided in geographical coordinates (EPSG code: 4326). Here we shall transform this original CRS (using function \texttt{transform\_coords}) to the projected North American Datum (NAD83, EPSG code: 5070).

\begin{verbatim}
dat_1 <- mk_track(dat_1, .x = x, .y = y, .t =t, crs = sp::CRS("+init=epsg:4326")) %>% 
  transform_coords(sp::CRS("+init=epsg:5070"))
\end{verbatim}

We then summarize the distribution of time intervals between successive locations to get a general impression for the sampling rate.

\begin{verbatim}
summarize_sampling_rate(dat_1)

## # A tibble: 1 x 9
##   min         q1               median mean   q3    max      sd     n unit 
##   <S3: table> <S3: table>      <S3: > <S3: > <S3:> <S3:> <dbl> <int> <chr>
## 1 0.1         1.93333333333333 2.033… 8.041… 2.56… 1208…  44.0  8957 min
\end{verbatim}

We see that we have 8957 total locations, the shortest interval between locations is 0.1 minutes and the largest time interval between locations is 1208 minutes, with median interval length equal to roughly 2 min. Despite the 2 min temporal resolution, we choose to resample the track to 10 min with a tolerance of 1 min (\texttt{track\_resample}), in order to conduct the analyses on the same temporal scale as the next example. The function \texttt{minutes} from the package \texttt{lubridate} \citep{lubridate}, is used here to create an object of class \texttt{Period} that is then passed to \texttt{track\_resample}. \texttt{Period}s can be specified using all common time units, thus it is straightforward to specify sampling rate and an acceptable tolerance. We will also choose to keep only those bursts (subsets of the track with constant sampling rate, within the specified tolerance) with at least three relocations, the minimum required to calculate a turn angle (\texttt{filter\_min\_n\_burst}). The following code implements those choices, and translates from a point representation to a step (step length, turn angle) representation of the data. In the final line of the code snippet, we use the function \texttt{time\_of\_day} \citep[a wrapper around \texttt{maptools::sunriset} and \texttt{maptools::crepuscule};][]{maptools} to calculate if a location was taken during the day or night. If the argument \texttt{include.crepuscule} is set to \texttt{TRUE}, the function not only considers day and night, but also dawn and dusk.

\begin{verbatim}
stps <- track_resample(dat_1, rate = minutes(10), tolerance = seconds(60)) %>% 
  filter_min_n_burst(min_n = 3) %>% steps_by_burst() %>% 
  time_of_day(include.crepuscule = FALSE)
\end{verbatim}

We then use the \texttt{str} function to inspect the structure of \texttt{stps}.

\begin{verbatim}
str(stps)
Classes ‘steps’, ‘tbl_df’, ‘tbl’ and 'data.frame':	1501 obs. of  11 variables:
 $ burst_  : num  16 16 16 16 18 18 21 21 23 23 ...
 $ x1_     : num  1780410 1780469 1780468 1780480 1780471 ...
 $ x2_     : num  1780469 1780468 1780480 1780469 1780469 ...
 $ y1_     : num  2412415 2412224 2412230 2412213 2412223 ...
 $ y2_     : num  2412224 2412230 2412213 2412217 2412217 ...
 $ sl_     : num  199.47 6.01 20.8 11.15 6.85 ...
 $ ta_     : num  NA 2.96 -2.7 -2.56 NA ...
 $ t1_     : POSIXct, format: "2010-02-10 03:00:36" "2010-02-10 03:10:40" ...
 $ t2_     : POSIXct, format: "2010-02-10 03:10:40" "2010-02-10 03:20:10" ...
 $ dt_     :Class 'difftime'  atomic [1:1501] 10.1 9.5 10.2 10.4 9.5 ...
 $ tod_end_: chr  "night" "night" "night" "night" ...

\end{verbatim}

\texttt{stps} is a regular \texttt{data\_frame} with 11 attributes of steps (e.g., start, end, and step length; columns) and 1501 steps (rows). For each step, the start (\texttt{x1\_}, \texttt{y1\_}) and end (\texttt{x2\_}, \texttt{y2\_}) coordinates, as well as the start and end time (\texttt{t1\_}, \texttt{t2\_}) are given. In addition the following derived quantities are calculated: step length (\texttt{sl\_}; in CRS units), turning angles (\texttt{ta\_}; in degrees; notice that it cannot be calculated for steps that are not preceded by a valid step), the time difference (\texttt{dt\_}), and the burst (\texttt{burst\_}) to which the step belongs. We proceed by preparing the environmental data. We hypothesized that Ricky T prefers forested wetlands over other landuse classes. We used the National Landcover Database (which is freely available at \url{https://www.mrlc.gov/nlcd11_data.php}). We first load the landuse raster and create a layer called wet that is 1 for forested wetlands (category 90) and 0 otherwise \citep[using the \texttt{raster} package;][]{raster}.

\begin{verbatim}
land_use <- raster("data/landuse_study_area.tif")
wet <- land_use == 90
\end{verbatim}

For convenience and readability, we give the layer a meaningful name.

\begin{verbatim}
names(wet) <- "wet"
\end{verbatim}

Before proceeding to modeling space use and habitat selection of Ricky T, we perform some exploratory data analysis based on step length and turning angles in different habitat types (wet forest versus other areas) and time of the day (day and night). We will have to extract the covariate value at the start point of each step (using the function \texttt{extract\_covariates}) and plot the density of step lengths per habitat class and time of day (Fig. 1; for the full code to replicate Fig. 1 see Supplement S1). Note that the function \texttt{extract\_covariates} takes an argument \texttt{where}, that indicates whether covariate values should be extracted at the beginning or the end of a step (\texttt{both} is also possible to extract the covariate at the start and the end of a step). Depending on the target process under investigation (habitat selection or movement), covariates might be extracted at the end of the step (habitat selection process) or at the start of the step (movement process). If covariates are extracted at the end of the step, they are typically included in the model as main effects, to answer questions of the type: How do covariates influence where the animal moves? In contrary, if covariates are extracted at the beginning of the step, they are typically included in the model as an interaction with movement characteristics (step length, log of the step length, or the cosine of the turn angle), to test hypotheses of the type: Do animals move faster/more directed, if they start in a given habitat? Finally, covariate values at the start and the end of a step can also be included in the model as interaction with each other, to test hypotheses of the type: Are animals more likely to stay in a given habitat, if they are already in the habitat?

In order to fit SSFs, the observed covariates associated with observed steps are compared to covariates associated with random steps. Random steps can be generated by either 1) sampling from the observed turn step-length and turn-angle distribution (resulting in a traditional SSF), or 2) by fitting a parametric distribution to the observed step lengths and turn angles (which can result in an iSSF). As mentioned above, an iSSF is arguably less biased, and also provides the user with a mechanistic movement model that can be used to simulate space use, and hence utilization distributions \citep{avgar2016, signer2017}. For these reasons, \texttt{amt} only implements the iSSFs with parametric distributions. For further details we refer the reader to \citet{duchesne2015equivalence} and \citet{avgar2016}.

Thus, we proceed by fitting a gamma distribution to the step lengths and a von Mises distribution to the turn angles using maximum likelihood \citep{circular,fitdistrplus}, and use these distributions to generate and pair 9 random steps with each observed step. Zero step lengths can cause estimation problems, so \texttt{random\_steps} automatically adds a random error between 0 and an upper limit that can be specified though the argument \texttt{random\_error} (with default= 0.001, alternatively the mimimum of observed step lengths would be a good choice). We then extract the covariates at the end point of each step (observed and random) using the function \texttt{extract\_covariates}, and fit a conditional logistic regression model to the resulting data including movement-related covariates with the function \texttt{fit\_issf} \citep[a wrapper to \texttt{survival::clogit};][]{survival-book}.

We included two main effects in the model, the environmental covariate \texttt{wet}, and the log of the step length (\texttt{log\_sl\_}) as a modifier of the shape parameter of the underlying gamma distribution. We also include interactions  between \texttt{wet} and \texttt{tod\_}, a factor with two levels -- day (the reference category) and night, and between \texttt{tod\_} and \texttt{log\_sl\_}. These interactions are included to the test the hypotheses that habitat selection and displacement rate, respectively, differ between day and night.

\begin{verbatim}
m1 <-stps %>% random_steps(n = 9) %>% extract_covariates(wet) %>% 
  time_of_day(include.crepuscule = FALSE) %>% 
  mutate(log_sl_ = log(sl_)) %>% 
  fit_issf(case_ ~ wet + log_sl_ + wet:tod_end_ + log_sl_:tod_end_ + 
		    strata(step_id_))
\end{verbatim}

We could have also included cosines of the turning angles and their interaction with day.  This choice would modify the concentration parameter of the underlying von Mises distribution for the turning angles and allow the degree of directional persistence to depend on time of day; the data summarized in Fig. 1 suggests that this could be a sensible choice. For the sake of simplicity, however, we have assumed we have correctly modeled the degree of directional persistence and that it does not differ between day and night.

Inspecting the fitted model (Table 1), we make the following observations. 1) there is evidence to suggest that the animal prefers forested wetlands over other landuse classes, 2) there is no difference in habitat preference between day and night, 3) there is evidence to modify the shape of the underlying gamma distribution (through the log of the step length), and 4) the modification of the shape parameter should be done separately for day and night.

Besides inspecting the coefficients and their standard errors, we can calculate derived quantities, such as the expected speed. Because we included an interaction between parameters of the step length distribution and time of the day, we have to account for this interaction when calculating the expected speed for day and night. We begin by retrieving the tentative parameter estimates for the gamma distribution of the step length distribution:

\begin{verbatim}
shape <- sl_shape(m1)
scale <- sl_scale(m1)
\end{verbatim}

And adjust the shape for day and night with the estimates of the corresponding coefficients from the fitted model \citep{avgar2016}.

\begin{verbatim}
shape_adj_day <- adjust_shape(shape, coef(m1)["log_sl_"])
shape_adj_night <- adjust_shape(shape, coef(m1)["log_sl_"]) + 
		      coef(m1)["log_sl_:tod_end_night"]
\end{verbatim}

The underlying gamma distributions for the step lengths vary by time of day (Table 1). The expected speed for day and night is then given by the product of the tentative scale parameter (no adjustment is needed here, because we did not include step length in the model) and the adjusted shape parameter. To obtain 95\% confidence intervals for the mean speed, we bootstrapped the model \texttt{m1} 1000 times by resampling (with replacement) the strata (for full code see Supplement S1). Results suggest that Ricky T moves significantly faster during nights (11.0 m/min, 95\% CI = 10.7, 11.4 m/min) than during days (8.57 m/min, 95\% CI = 7.8, 9.32 m/min).

In a final step, we simulated space-use from the fitted model \texttt{m1} to obtain a model-based estimate of the animal's utilization distribution \citep[UD;][]{avgar2016, signer2017}. Generally, two types of UDs can be simulated: the transient UD and the steady state UD. The transient UD describes the expected space-use distribution of the animal within a short time period, and is hence conditional on the starting position. The steady state UD describes the expected space-use distribution of the animal in the long-term. In order to simulate UDs one has to ensure that the animals stay within the study domain. We see three possible methods for achieving this goal: 1) use a covariate that attracts the animal towards one or more centers of activity (e.g., the squared distance to the mean of all coordinates), 2) use a very large landscape, or 3) use a wrapped landscape (torus). Here, we illustrate the simulation of steady state and transient UDs. For the steady state UD we simulate from the first observed location $10^7$ time steps on a toroid landscape, once for day and once for night. For the transient UD, we are interested in the UD up to 10 hours after last observation, we therefore simulated 72 steps (at a 10 min sampling rate) $5\times 10^3$ times.

We describe the simulation for the steady state and transient UD for daytime. First we create a movement kernel (Fig. 2A), that is used to determine the animal's movement ability at each time step. Note, we use the tentative scale estimate and the shape estimate adjusted for day.

\begin{verbatim}
mk <- movement_kernel(scale = scale, shape = shape_adj_day, template = wet)
\end{verbatim}

Next, we create a habitat kernel (that is for each pixel we calculate the estimated selection coefficients times the resources and exponentiate the product; Fig. 2B).

\begin{verbatim}
hk <- habitat_kernel(coef = list(forest = coef(m1)["wet"]), resources = wet)
\end{verbatim}

We then estimate the steady state UD (Fig. 2CE) with the function \texttt{simulate\_ud}:

\begin{verbatim}
ssud_day <- simulate_ud(movement_kernel = mk, 
		        habitat_kernel = hk, 
		        start = as.numeric(stps[1, c("x1_", "y1_")]), 
		        n = 1e7)
\end{verbatim}

In order to simulate the transient UD (Fig. 2CE), we have to repeatedly simulate short tracks starting at the same point, and then sum individual UDs and normalize, which we do with a simple for-loop.

\begin{verbatim}
tud <- wet_c
tud[] <- 0
for(i in 1:5e3) {
  tud <- tud + simulate_ud(mk, hk, 
                           as.numeric(stps[1501, c("x1_", "y1_")]), 
                           n = 72)
}
tud[] <- tud[] / sum(tud[])
\end{verbatim}

All simulations took $< 1$ minutes on a standard laptop. 

\subsection*{Many animals: quantifying population-level effects}

We start again with the same data set (\texttt{dat}), containing data from 6 individual fishers. This time we are interested in quantifying among-animal variability in the selection coefficients. We proceed using nearly all the same steps as in the first example, but with a different data structure: \texttt{data\_frames} with list-columns \citep{tibble}. List columns are best thought of as regular columns of a \texttt{data\_frame} that are R lists and can contain any objects (in our case tracks and fitted models). The \texttt{purrr::nest} command can be used to nest data into a list-column \citep{purrr}.

\begin{verbatim}
dat_all <- dat %>% nest(-id) 
\end{verbatim}

\texttt{dat\_all} is now a \texttt{data\_frame} with 6 rows (one for each individual) and two columns. In the first column the animal id is given, and in the second column (by default named data) the relocations of the corresponding animal are saved. We start by assigning the sex of each animal.

\begin{verbatim}
dat_all$sex <- c("f", "f", "f", "m", "m", "m")
\end{verbatim}

We can now apply the steps as before for all animals.  We first create a track for each animal and transform the coordinate reference system.

\begin{verbatim}
dat_all <- dat_all %>% 
  mutate(trk = lapply(data, function(d) {
    mk_track(d, x, y, t, crs = sp::CRS("+init=epsg:4326")) %>% 
      transform_coords(sp::CRS("+init=epsg:5070"))
  }))
\end{verbatim}

Next, we prepare again the landuse data. This time we reclassify the landuse raster (using \texttt{raster::reclassify}) into five categories: water and wetland forests, developed open spaces, other developed areas, forests and shrubs, and crops.

\begin{verbatim}
land_use <- raster("data/landuse_study_area.tif")
rcl <- cbind(c(11, 12, 21:24, 31, 41:43, 51:52, 71:74, 81:82, 90, 95), 
             c(1, 1, 2, 3, 3, 3, 2, 4, 4, 4, 4, 4, 4, 4, 4, 4, 5, 5, 1, 1))
lu <- reclassify(land_use, rcl)
names(lu) <- "landuse"
\end{verbatim}

We again first inspect the sampling rate of the 6 individuals:

\begin{verbatim}
dat_all %>% mutate(sr = lapply(trk, summarize_sampling_rate)) %>% 
  select(id, sr) %>% unnest
  
# A tibble: 6 x 10
     id   min    q1 median  mean    q3   max    sd     n unit 
  <int> <dbl> <dbl>  <dbl> <dbl> <dbl> <dbl> <dbl> <int> <chr>
1  1072 3.92   9.80  10.0  20.7  10.3   1650  95.2  1348 min  
2  1465 0.400  1.97   2.03  8.93  3.98  1080  48.1  3003 min  
3  1466 0.417  1.97   2.07 15.7   4.08  2167 104    1500 min  
4  1078 1.35   9.78  10.0  21.7  10.3   2111  87.2  1637 min  
5  1469 0.417  1.97   2.17 13.3   5.42  2889  90.2  2435 min  
6  1016 0.100  1.93   2.03  8.04  2.57  1209  44.0  8957 min  
\end{verbatim}

This time some individuals have a 2 min sample rate and others a 10 min one. Thus we decided to resample the tracks to the same sampling rate of 10 minutes \citep[noting that (i)SSF inference is scale dependent;][]{signer2017} using \texttt{track\_resample}. We then filter again bursts, keeping only those with at least three points (\texttt{filter\_min\_n\_burst}), convert from a point to a step representation of the tracks (\texttt{steps\_by\_burst}), generate 9 random steps for each observed step  (\texttt{random\_steps}), extract the environmental covariates (\texttt{extract\_covariates}), convert landuse to a factor (\texttt{mutate}) and fit a SSF (\texttt{fit\_issf}). The main difference to the previous example here, is that the all the steps from above are wrapped into one \texttt{mutate} call. This call creates a new column to \texttt{dat\_all} called \texttt{ssf}. This is a list column and each entry in this columns contains a fitted SSF.

\begin{verbatim}
m1 <- dat_all %>% mutate(ssf = lapply(trk, function(x) {
  x %>% track_resample(rate = minutes(10), tolerance = minutes(2)) %>% 
    filter_min_n_burst(min_n = 3) %>% 
    steps_by_burst() %>% random_steps() %>% 
    extract_covariates(lu) %>% 
    mutate(landuse_end = factor(landuse)) %>% 
    fit_issf(case_ ~ landuse_end + strata(step_id_))
})) 
\end{verbatim}

\texttt{m1} is still a \texttt{data\_frame} with one new column: \texttt{ssf}, that is again a list column with a fitted SSF. From here, it is easy to investigate coefficients for several animals and look at population-level effects. The results suggest that there are some general population-level trends (Fig. 3). All fisher seem to prefer wetland forests and natural areas relative to developed areas (of either type), whereas considerable among-animal variability in the coefficients for crops makes it difficult to draw firm conclusions about this landuse type. Lastly, there seems to be little differentiation based on sex. (Fig. 3, code provided in Supplement S1).

\section*{Discussion and Outlook}
We have illustrated how \texttt{amt} can be used to fit Step-Selection Functions (SSFs) and explore temporal movement and habitat selection patterns at the individual and population levels. We demonstrated how an iSSF, fit to a single fisher, can be used to simulate utilization distributions \citep[UDs][]{signer2017}. The UD map (Fig. 2) can be thought as of a stochastic approximation of a mechanistic home-range model \citep{moorcroft2013mechanistic}. Whereas traditional home-range estimators offer static summaries of space-use patterns, mechanistic home-range estimators can provide insights into the movement and habitat selection processes that give rise to these patterns.  In our model, we included an interaction between parameters of the movement model and time of the day (day/night), allowing us to explore time-dependent space-use patterns (Fig. 2B,D). 
We then showed how \texttt{amt} can be used to conduct similar analyses with more than one animal, allowing us to investigate population-level effects by looking at the distribution of animal-specific coefficients \citep[Fig. 3][]{fieberg2010correlation}. In our example we restricted the analysis to habitat selection, incorporation of a movement model would be straight forward here as well \citep{prokopenko2017characterizing,scraffordpress}.

We expect \texttt{amt} will contribute to movement ecology in two ways. First, \texttt{amt} is likely to help researchers manage their data and analyses using  a more reproducible workflow, a much discussed issue \citep[e.g.,][]{lewis2018,bes2017}. 
Second, \texttt{amt} will facilitate the use of iSSFs by a wider community of ecologists and also allow them to more fully realize the power of these methods (e.g., by modeling how landscape features influence both movement and habitat selection processes). Prior to \texttt{amt}, software for implementing iSSFs was not available. Therefore, use of iSSFs required custom-written code. \texttt{amt} provides functions that make it easy to fit iSSFs and to explore predicted space-use patterns from fitted models.

Besides the introduced functions to fit SSFs, \texttt{amt} provides additional functions for calculating home ranges, estimating RSFs and other utility functions to work with telemetry data and interface with other packages. Future development of \texttt{amt} will focus on increased functionality by adding more functions for data quality assurance. We also hope to implement UD estimation for more sophisticated models. Currently, \texttt{amt} does not allow simulating UDs from models that include interactions between movement (speed and/or turning angles) and other covariates.

\newpage

\section*{Acknowledgement}
We thank Niko Balkenhol, Scott LaPoint, Laura Richter and Christoph Raab for comments on ealier versions of this manuscript.

\section*{Authors contribution}

JS, JF and TA conceived the ideas and designed the package; JS programmed the package; JS, JF and TA analysed the case study; JS, JF and TA wrote the manuscript. All authors contributed critically to the drafts and gave final approval for publication.

\section*{Figures}

\begin{figure}[ht!]
 \centering
 \includegraphics[width=\linewidth]{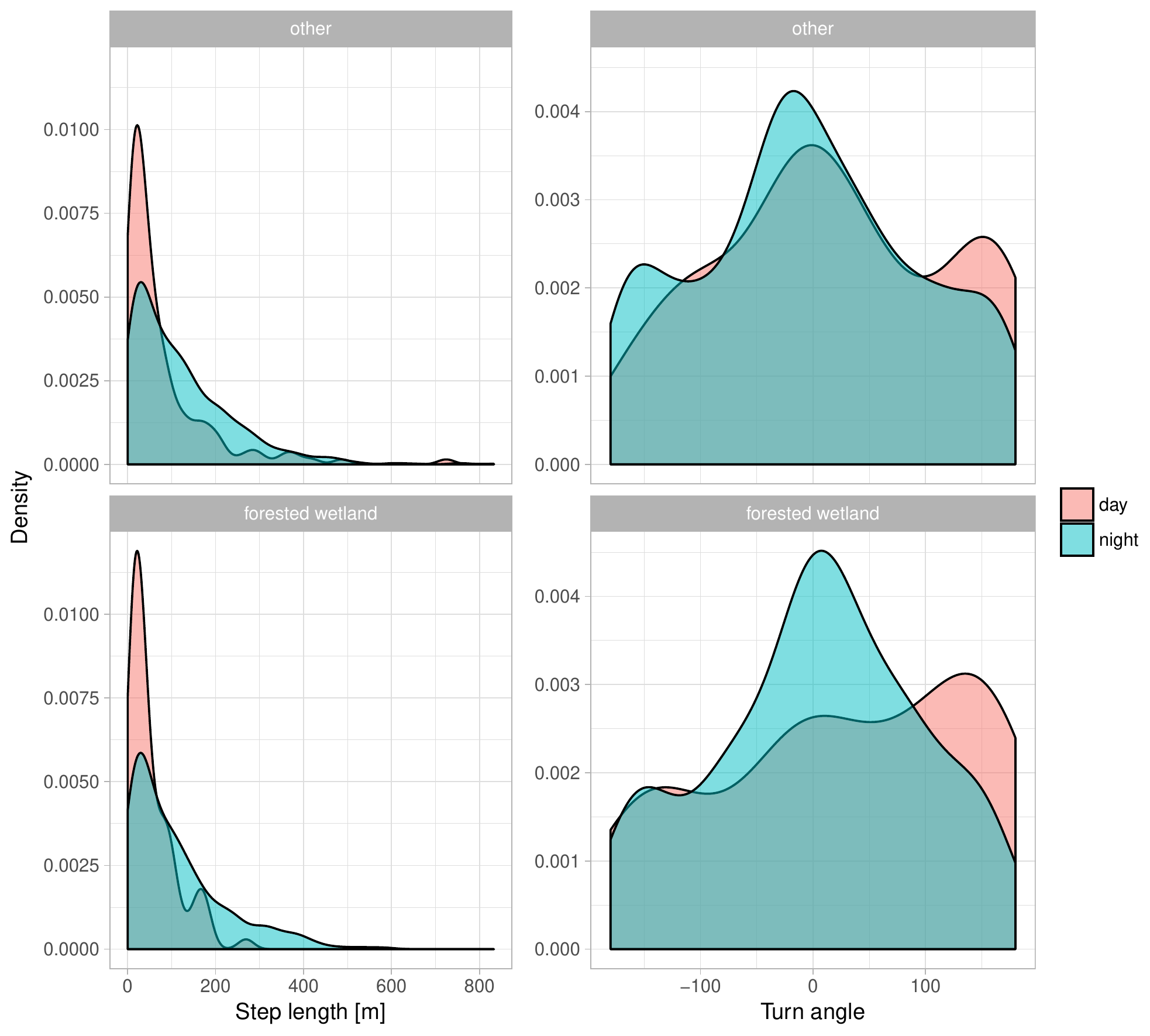}
 % fig_eda_1_animal.pdf: 0x0 pixel, 300dpi, 0.00x0.00 cm, bb=
 \label{fig:eda}
 \caption{Exploratory data analysis of one individual fisher, Ricky T (id: 1016): empirical distributions of step lengths (first column) and turning angles (second column) are shown for forested wetland (second row) and other habitats (first row) and for day and night (colors).}
\end{figure}

\begin{figure}[ht!]
 \centering
 \includegraphics[width=1\textwidth]{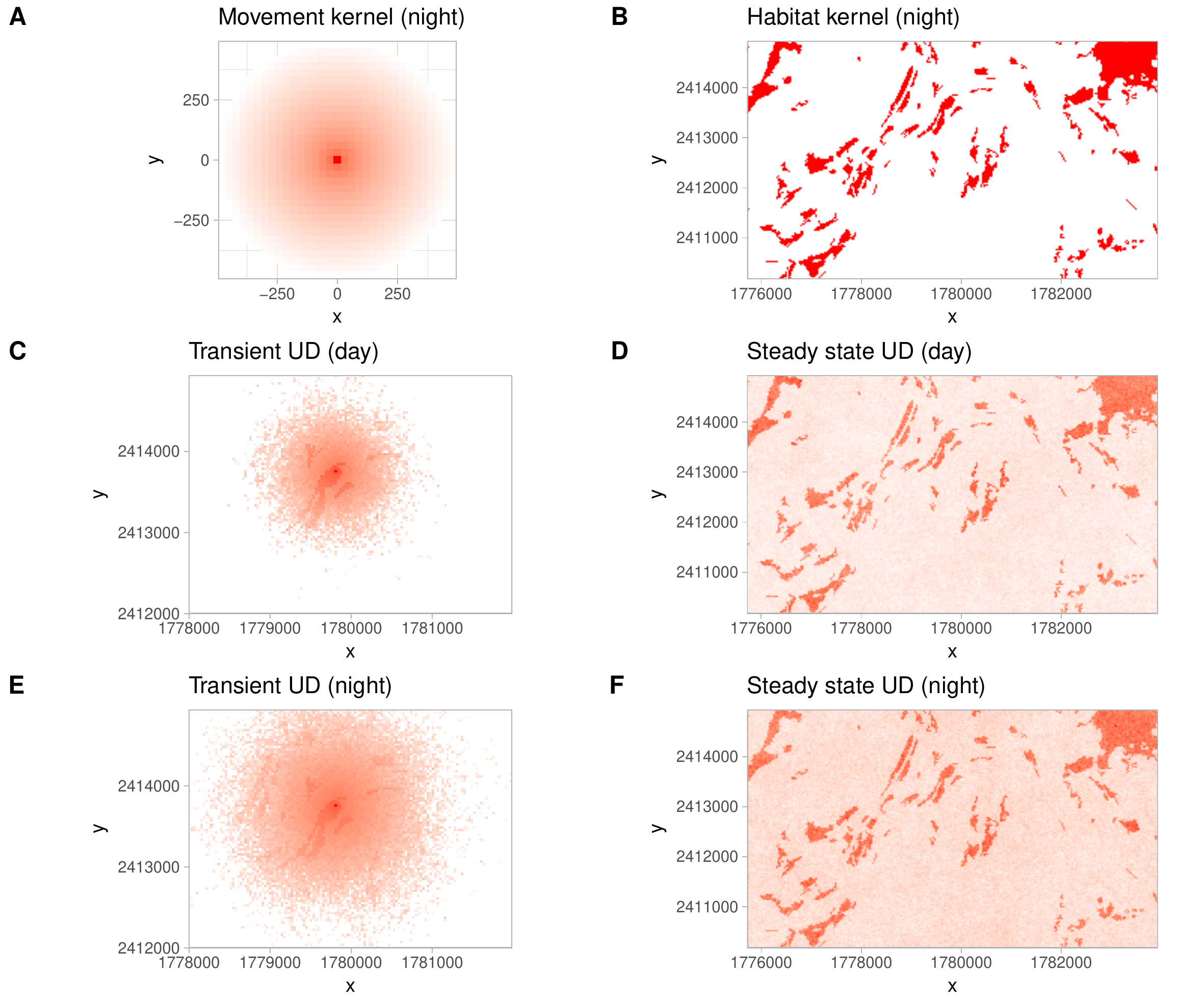}
 % fig_one_animal1.pdf: 0x0 pixel, 300dpi, 0.00x0.00 cm, bb=
 \label{fig:oneanimal}
 \caption{Simulated utilization distributions. To obtain simulated Utilization Distributions (UD), a movement kernel (panel A) and a habitat kernel (panel B) are needed. The movement kernel is always placed at the current position of the animal. The next step of the animal is then sampled with probability proportional to the product of two kernels. Expected differences in movement speeds between night and day are reflected in the transient UD (panels C and E) and to a lesser extend in steady state UD (panels D and F). Note, for better visualization, fills were log10 transformed for panels A, C, and E.}
 \end{figure}

\begin{figure}[ht!]
 \centering
 \includegraphics[width=1.1\textwidth]{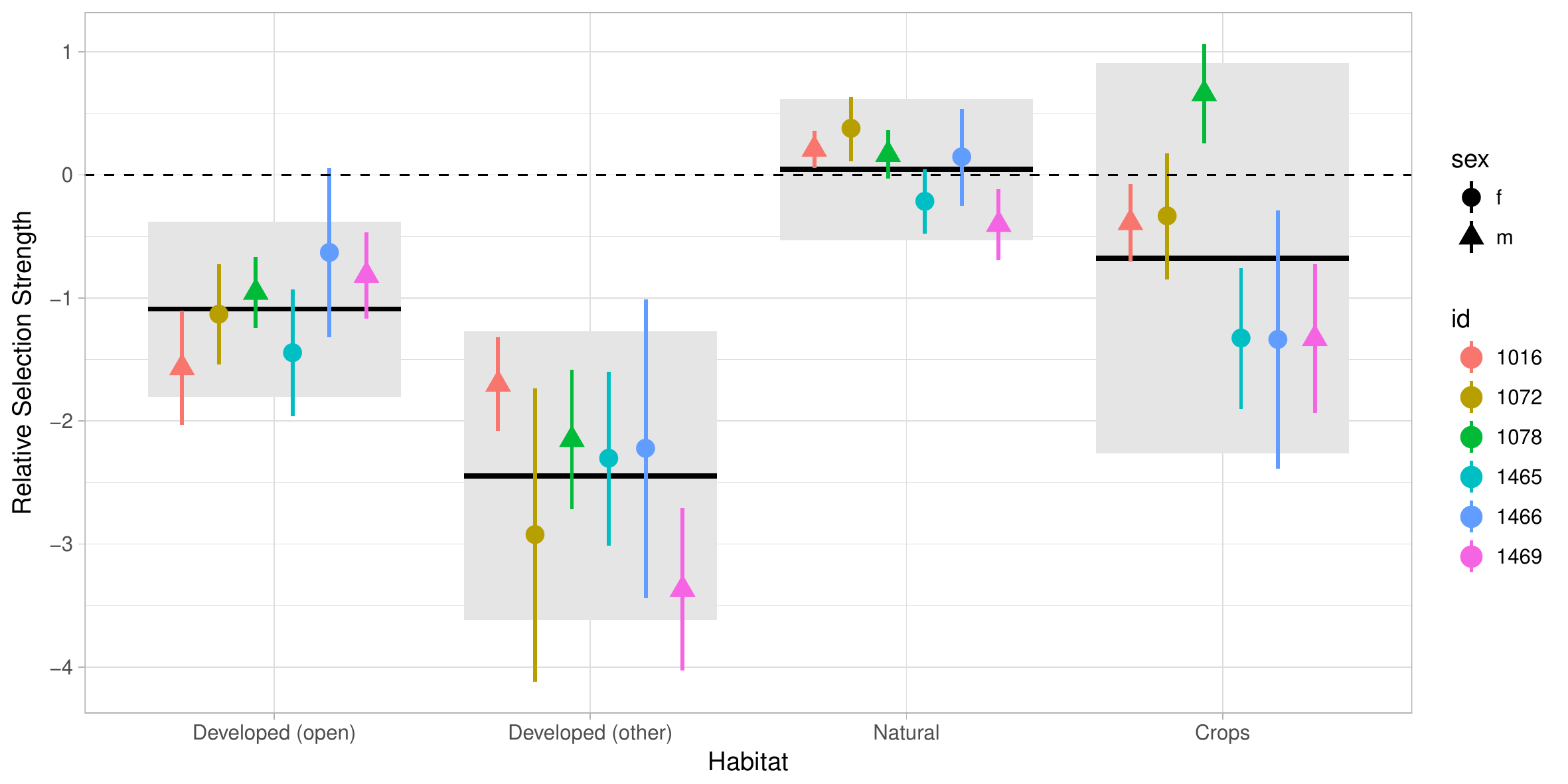}
 % fig_all_animals.pdf: 0x0 pixel, 300dpi, 0.00x0.00 cm, bb=
 \label{fig:allanmials}
 \caption{Point estimates with 95\% confidence intervals for the relative selection strength \citep{avgar2017relative} for different landuse classes (we used wetland forests and wet areas as the reference class). Different colors indicate the id of the animals and symbols the sex (circles for female and triangles for males). Population-level estimates are given by solid vertical lines and 95\% confidence intervals at population level is given by the light gray boxes. The dashed horizontal line indicates no preference relative to wetland forest (the reference category).}
\end{figure}

\newpage

\section*{Tables}

Table 1: Coefficients of fitted integrated Step Selection Function.

\begin{table}[ht]
\centering
\begin{tabular}{rrrrrr}
  \hline
 & coef & exp(coef) & se(coef) & z & Pr($>$$|$z$|$) \\ 
  \hline
wet & 0.9765 & 2.6552 & 0.2672 & 3.6551 & 0.0003 \\ 
  log\_sl\_ & -0.2775 & 0.7577 & 0.0600 & -4.6259 & 0.0000 \\ 
  wet:tod\_end\_night & -0.3656 & 0.6938 & 0.2831 & -1.2914 & 0.1966 \\ 
  log\_sl\_:tod\_end\_night & 0.3529 & 1.4231 & 0.0655 & 5.3839 & 0.0000 \\ 
   \hline
\end{tabular}
\end{table}

\newpage

\bibliography{lib}{}

\newpage

\section*{Supplement S1}
Source code to reproduce the two examples.

\subsection*{Example 1: One animal}
\begin{verbatim}
library(raster)
library(lubridate)
library(amt) 
set.seed(123)

dat <- read_csv("data/Martes pennanti LaPoint New York.csv") %>% 
  filter(!is.na(`location-lat`)) %>% 
  select(x = `location-long`, y = `location-lat`, 
         t = `timestamp`, id = `tag-local-identifier`) %>% 
  filter(id %in% c(1465, 1466, 1072, 1078, 1016, 1469)) # for example 2
dat_1 <- dat %>% filter(id == 1016)


dat_1 <- mk_track(dat_1, x, y, t, crs = sp::CRS("+init=epsg:4326")) %>% 
  transform_coords(sp::CRS("+init=epsg:5070"))

summarize_sampling_rate(dat_1)

stps <- track_resample(dat_1, rate = minutes(10), tolerance = minutes(1)) %>%
  filter_min_n_burst(min_n = 3) %>% steps_by_burst() %>%
  time_of_day(include.crepuscule = FALSE)
str(stps, width = 80, strict.width = "no", nchar.max = 80, give.attr = FALSE)



land_use <- raster("data/landuse_study_area.tif")
wet <- land_use == 90
names(wet) <- "wet"

eda1 <- stps %>% extract_covariates(wet, where = "start") %>% 
  mutate(landuse = factor(wet, levels = c(0, 1), labels = c("other", "forested wetland"))) 

# Plot 1 ------------------------------------------------------------------
  
p1 <- eda1 %>% select(landuse, tod = tod_end_, sl_, ta_) %>% 
  gather(key, val, -landuse, -tod) %>% 
  filter(key == "sl_") %>% 
  ggplot(., aes(val, group = tod, fill = tod)) + geom_density(alpha = 0.5) +
  facet_wrap(~ landuse, nrow = 2) +
  xlab("Step length [m]") + theme_light() +
  ylab("Density") +
  theme(legend.title = element_blank())

p2 <- eda1 %>% select(landuse, tod = tod_end_, sl_, ta_) %>% 
  gather(key, val, -landuse, -tod) %>% 
  filter(key == "ta_") %>% 
  ggplot(., aes(val, group = tod, fill = tod)) + geom_density(alpha = 0.5) +
  facet_wrap(~ landuse, nrow = 2) +
  xlab("Turn angle") + theme_light() +
  theme(legend.title = element_blank(), 
        axis.title.y = element_blank())

library(cowplot)
pg1 <- plot_grid(
  p1 + theme(legend.position = "none"), 
  p2 + theme(legend.position = "none"), rel_widths = c(1, 1)
)

leg <- get_legend(p1)
plot_grid(pg1, leg, rel_widths = c(1, 0.1))

ggsave("fig_eda_1_animal.pdf", width = 20, height = 18, units = "cm")

#####################################

# fit the model
m1 <-stps %>% random_steps(n = 9) %>% extract_covariates(wet) %>% 
  time_of_day(include.crepuscule = FALSE) %>% 
  mutate(log_sl_ = log(sl_)) -> d1

m1 <- d1 %>% fit_issf(case_ ~ wet +  sl_  + wet:tod_end_+  sl_:tod_end_ + strata(step_id_))
m1 <- d1 %>% fit_issf(case_ ~ wet + log_sl_ + wet:tod_end_+ log_sl_:tod_end_ + strata(step_id_))
m1 <- d1 %>% fit_issf(case_ ~ wet + log_sl_ + sl_  + wet:tod_end_+ log_sl_:tod_end_ + sl_:tod_end_ + strata(step_id_))

AIC(m1$model)
summary(m1)

s <- summary(m1$model)$coefficients
s
print(xtable::xtable(s, digits = 4,
                     type = "latex", 
                     caption.placement = "top"))

shape <- sl_shape(m1)
scale <- sl_scale(m1)

shape_adj_day <- adjust_shape(shape, coef(m1)["log_sl_"])
shape_adj_night <- adjust_shape(shape, coef(m1)["log_sl_"]) + 
  coef(m1)["log_sl_:tod_end_night"]

scale_adj_day <- adjust_scale(scale, coef(m1)["sl_"])
scale_adj_night <- adjust_scale(scale, coef(m1)["sl_"]) + 
  coef(m1)["sl_:tod_end_night"]

# speed
speed_day <- shape * scale_adj_day
speed_night <- shape * scale_adj_night

speed_day <- shape_adj_day * scale
speed_night <- shape_adj_night * scale

speed_day <- shape_adj_day * scale_adj_day
speed_night <- shape_adj_night * scale_adj_night

scale
shape

shape_adj_day
shape_adj_night

x <- seq(1, 500, 1)
plot(x, dgamma(x, shape = shape_adj_night, scale = scale_adj_night), type = "l")
lines(x, dgamma(x, shape = shape_adj_day, scale = scale_adj_day), type = "l", add = TRUE)


# Boot everything
mod_data <- stps %>% random_steps(n = 9) %>% 
  extract_covariates(wet, where = "end") %>% 
  time_of_day(include.crepuscule = FALSE) %>% 
  mutate(log_sl_ = log(sl_), cos_ta_ = cos(as_rad(ta_)))

strata <- unique(mod_data$step_id_)
n <- length(strata)

bt <- replicate(1000, {
  m_boot <- mod_data[mod_data$step_id_ %in% sample(strata, n, TRUE), ] %>% 
    fit_clogit(case_ ~ wet + log_sl_ + sl_ + wet:tod_end_ + 
                 sl_:tod_end_ + log_sl_:tod_end_ + strata(step_id_))
  
  scale_adj_day <- adjust_scale(shape, coef(m_boot)["sl_"])
  scale_adj_night <- adjust_scale(shape, coef(m_boot)["sl_"]) + 
    coef(m_boot)["sl_:tod_end_night"]
  
  shape_adj_day <- adjust_shape(shape, coef(m_boot)["log_sl_"])
  shape_adj_night <- adjust_shape(shape, coef(m_boot)["log_sl_"]) + 
    coef(m_boot)["log_sl_:tod_end_night"]
  
  ## speed
  c(shape_adj_day * scale_adj_day,
    shape_adj_night * scale_adj_night)
})

bt2 <- data_frame(
  rep = 1:ncol(bt),
  day = bt[1, ], 
  night = bt[2, ]
) %>% gather(key, val, -rep) 


bt2 %>% group_by(key) %>% summarise(lq = quantile(val, 0.025), 
                                    me = median(val), 
                                    mean = mean(val), 
                                    uq = quantile(val, 0.975))
  
# m/min 
bt2 %>% group_by(key) %>% summarise(lq = quantile(val, 0.025) / 10, 
                                    me = median(val) / 10,  
                                    mean = mean(val) / 10, 
                                    uq = quantile(val, 0.975) / 10)

## Simulate ud
wet_c <- crop(wet, amt::bbox(dat_1, spatial = TRUE, buff = 1e3))

mk <- movement_kernel(scale, shape_adj_day, wet_c)
hk <- habitat_kernel(list(wet = coef(m1)["wet"]), wet_c)


system.time(ssud_day <- simulate_ud(mk, hk, 
                                    as.numeric(stps[1, c("x1_", "y1_")]), 
                                    n = 1e7))
plot(ssud_day)

tud <- wet_c
tud[] <- 0
system.time(for(i in 1:5e3) tud <- tud + 
              simulate_ud(mk, hk, as.numeric(stps[1501, c("x1_", "y1_")]), n = 72))
tud[] <- tud[] / sum(tud[])
tud_day <- tud
plot(tud_day)

# night
mk <- movement_kernel(scale, shape_adj_night, wet_c)
hk <- habitat_kernel(list(wet = coef(m1)["wet"] + 
                            coef(m1)["wet:tod_end_night"]), wet_c)

system.time(ssud_night <- simulate_ud(
  mk, hk, as.numeric(stps[1, c("x1_", "y1_")]), n = 1e7))
plot(ssud_night)

tud <- wet_c
tud[] <- 0
system.time(for(i in 1:5e3) tud <- tud + 
              simulate_ud(mk, hk, as.numeric(stps[1501, c("x1_", "y1_")]), n = 72))
tud[] <- tud[] / sum(tud[])
tud_night <- tud
plot(tud1 <- crop(tud_day, extent(c(1778000, 1782000, 2412000, 2415000))))
plot(tud2 <- crop(tud_night, extent(c(1778000, 1782000, 2412000, 2415000))))


pllog <- list(
  geom_raster(),
  coord_equal(),
  scale_fill_continuous(low = "white", high = "red", 
                        tran = "log10", na.value = "white"),
  scale_y_continuous(expand = c(0, 0)),
  scale_x_continuous(expand = c(0, 0)),
  theme_light(),
  theme(legend.position = "none"))

pl <- list(
  geom_raster(),
  coord_equal(),
  scale_fill_continuous(low = "white", high = "red", na.value = "white"),
  scale_y_continuous(expand = c(0, 0)),
  scale_x_continuous(expand = c(0, 0)),
  theme_light(),
  theme(legend.position = "none"))

r1 <- data.frame(rasterToPoints(mk))
p1 <- ggplot(r1, aes(x, y, fill = d)) + pllog + ggtitle("Movement kernel (night)")

r2 <- data.frame(rasterToPoints(hk))
p2 <- ggplot(r2, aes(x, y, fill = layer)) +  pl + ggtitle("Habitat kernel (night)")


r1 <- data.frame(rasterToPoints(tud1))
p3 <- ggplot(r1, aes(x, y, fill = layer)) + pllog + ggtitle("Transient UD (day)")

r2 <- data.frame(rasterToPoints(tud2))
p4 <- ggplot(r2, aes(x, y, fill = layer)) +  pllog + ggtitle("Transient UD (night)")


r1 <- data.frame(rasterToPoints(ssud_day))
p5 <- ggplot(r1, aes(x, y, fill = layer)) + pl + ggtitle("Steady state UD (day)")

r2 <- data.frame(rasterToPoints(ssud_night))
p6 <- ggplot(r2, aes(x, y, fill = layer)) +  pl + ggtitle("Steady state UD (night)")

plot_grid(p1, p2, p3, p5, p4, p6, ncol = 2, labels = "AUTO")



ggsave("img/fig_one_animal1.pdf", height = 20, width = 24, units = "cm")
\end{verbatim}

\subsection*{Example 2: Many animals}
\begin{verbatim}
library(ggplot2)
library(raster)
library(amt) 
library(lubridate)
library(amt) 
library(parallel)

dat <- read_csv("data/Martes pennanti LaPoint New York.csv") %>% 
  filter(!is.na(`location-lat`)) %>% 
  select(x = `location-long`, y = `location-lat`, 
         t = `timestamp`, id = `tag-local-identifier`) %>% 
  filter(id %in% c(1465, 1466, 1072, 1078, 1016, 1469))

dat_all <- dat %>% nest(-id) 
dat_all$sex <- c("f", "f", "f", "m", "m", "m")
dat_all <- dat_all %>% 
  mutate(trk = map(data, function(d) {
    mk_track(d, x, y, t, crs = sp::CRS("+init=epsg:4326")) %>% 
      transform_coords(sp::CRS("+init=epsg:5070"))
  }))
dat_all %>% mutate(sr = lapply(trk, summarize_sampling_rate)) %>% 
  select(id, sr) %>% unnest

land_use <- raster("data/landuse_study_area.tif")
rcl <- cbind(c(11, 12, 21:24, 31, 41:43, 51:52, 71:74, 81:82, 90, 95), 
             c(1, 1, 2, 3, 3, 3, 2, 5, 5, 5, 5, 5, 5, 5, 5, 5, 8, 8, 1, 1))
# water, dev open, dev, barren, forest, shrub and herb, crops, wetlands
# 1: water, wetlands
# 2: developed (open)
# 3: developed (other)
# 5: forest, herbaceouse
# 8: crops
lu <- reclassify(land_use, rcl, right = NA)
names(lu) <- "landuse"

m1 <- dat_all %>% 
  mutate(steps = map(trk, function(x) {
    x %>% track_resample(rate = minutes(10), tolerance = seconds(120)) %>% 
      filter_min_n_burst() %>% 
      steps_by_burst() %>% random_steps() %>% 
      extract_covariates(lu, where = "both") %>% 
      mutate(landuse_end = factor(landuse_end)) 
    })) 


m1 <- m1 %>% mutate(fit = map(steps, ~ fit_issf(., case_ ~ landuse_end + 
                                             strata(step_id_))))

    
d2 <- m1 %>% mutate(coef = map(fit, ~ broom::tidy(.x$model))) %>% 
  select(id, sex, coef) %>% unnest %>% 
  mutate(id = factor(id)) %>% group_by(term) %>% 
  summarize(
    mean = mean(estimate), 
    ymin = mean - 1.96 * sd(estimate), 
    ymax = mean + 1.96 * sd(estimate)
  )

d2$x <- 1:nrow(d2)


p1 <- m1 %>% mutate(coef = map(fit, ~ broom::tidy(.x$model))) %>% 
  select(id, sex, coef) %>% unnest %>% 
  mutate(id = factor(id)) %>% 
  ggplot(., aes(x = term, y = estimate, group = id, col = id, pch = sex)) +
  geom_rect(mapping = aes(xmin = x - .4, xmax = x + .4, ymin = ymin, 
                          ymax = ymax), data = d2, inherit.aes = FALSE,
                  fill = "grey90") +
  geom_segment(mapping = aes(x = x - .4, xend = x + .4, 
                             y = mean, yend = mean), data = d2, inherit.aes = FALSE,
                  size = 1) +
  geom_pointrange(aes(ymin = conf.low, ymax = conf.high), 
                  position = position_dodge(width = 0.7), size = 0.8)  +
  geom_hline(yintercept = 0, lty = 2) +
  labs(x = "Habitat", y = "Relative Selection Strength") +
  theme_light() + 
  scale_x_discrete(labels = c("Developed (open)", "Developed (other)", "Natural", "Crops")) 

p1

ggsave("img/fig_all_animals.pdf", width = 24, height = 12, units = "cm")
\end{verbatim}

\end{document}